# Three-Dimensional Rigid-Body Impact Mechanics for Automobile Collisions

Micky C. Marine

## Abstract

Two-dimensional (planar) rigid-body impact mechanics for application in automobile collisions have been described by a number of researchers over the last several decades. Little has been discussed, however, regarding three-dimensional rigid-body impact mechanics in this regard. Two commercially available accident simulation programs, PC-Crash and Virtual CRASH, offer three-dimensional rigid-body impact mechanics as one of their collision models but documentation of the complete development of their three-dimensional equations, particularly with respect to necessary constraint strategies at the impulse center, are not readily available. In this paper, a three-dimensional rigid-body impact mechanics derivation is presented. In order to solve the set of impact mechanics equations of motion it is necessary to develop constraint relationships. The constraint strategy described in the literature pertaining to the PC-Crash Full-Impact/Sliding-Impact scenarios for two dimensions is extended to the three-dimensional case and the ramifications regarding post-impact relative velocity at the impulse center is discussed. A second strategy in which an impulse component is aligned with the contact plane component of the initial relative velocity at the impulse center is presented and compared to the PC-Crash scenarios. Lastly, while these two strategies incorporate a single impulse ratio/friction parameter for the contact plane, a strategy involving two independent impulse ratio/friction parameters is briefly discussed.

## Introduction

Over the last five decades much has been written regarding the application of planar rigid-body impact mechanics (PIM) to automobile collisions. Emori [1], Woolley [2, 3], Limpert [4], Ishikawa [5, 6], Smith [7], and Brach [8-11] have all presented PIM analyses, with the usage of various impulse center constraints, for automobile collision application. While we have numerous references to rely on for PIM analyses, little has been written on three-dimensional rigid-body impact mechanics (3DIM) applied to automobile collisions. The commercially available accident simulation programs PC-Crash and Virtual CRASH both incorporate a 3DIM model as one of their collision simulation models. Though these programs have been commercially available for some time, technical literature regarding the relevant contact plane constraint assumptions underlying their 3DIM models is scarce. Both the PC-Crash Technical Manual [12] and the Virtual CRASH User's Guide [13] provide derivations for two-dimensional PIM configurations. When developing a 3DIM formulation, important decisions must be made regarding the application of impulse center constraints along the plane perpendicular to the defined normal axis (i.e., contact plane). In this paper we will examine 3DIM contact plane configuration strategies and look at the ramifications that arise in their use regarding relative velocities at the impulse center.

The development of the salient three-dimensional 3DIM equations begins with the Newton-Euler equations of rigid-body motion:

$$\vec{F} = m\frac{d\vec{V}}{dt}$$

$$\vec{M} = \underline{I}\frac{d\vec{\omega}}{dt} + \vec{\omega} \times \underline{I} \cdot \vec{\omega} \tag{1}$$

Applying the assumptions that the only significant impulses acting on the rigid body are due to contact forces between bodies; no internal moments are caused by these forces; and that the impulse occurs instantaneously with no translational or rotational displacement of the rigid body (thus, impulses from compliant sources, e.g., like tire/terrain forces, are not included), equations 1 can be rewritten as:

$$\vec{P}_1 = \int \vec{F}_1 dt = m_1 \Delta \vec{V}_1 \qquad \vec{P}_2 = \int \vec{F}_2 dt = m_2 \Delta \vec{V}_2$$

$$\vec{r}_1 \times \vec{P}_1 = \vec{r}_1 \times \int \vec{F}_1 dt = \underline{I}_1 \Delta \vec{\omega}_1 \qquad \vec{r}_2 \times \vec{P}_2 = \vec{r}_2 \times \int \vec{F}_2 dt = \underline{I}_2 \Delta \vec{\omega}_2 \tag{2}$$

Where, in a vehicle-based coordinate system, the impulses and impulse center position vectors relative to the vehicle centers of gravity (CGs) are expressed in equations 3:

$$\vec{P}_1 = [P_{x1} \quad P_{y1} \quad P_{z1}]^T \qquad \vec{P}_2 = [P_{x2} \quad P_{y2} \quad P_{z2}]^T \tag{3}$$

$$\vec{r}_1 = [x_1 \quad y_1 \quad z_1]^T \qquad \vec{r}_2 = [x_2 \quad y_2 \quad z_2]^T$$

Using these equations, the change in velocity at the impulse center for each vehicle is found to be:

$$\vec{V}_{IC1f} - \vec{V}_{IC1i} = \Delta \vec{V}_1 + \Delta \vec{\omega}_1 \times \vec{r}_1 = \frac{1}{m_1}\vec{P}_1 + [\underline{I}_1^{-1}(\vec{r}_1 \times \vec{P}_1)] \times \vec{r}_1$$

$$\vec{V}_{IC2f} - \vec{V}_{IC2i} = \Delta \vec{V}_2 + \Delta \vec{\omega}_2 \times \vec{r}_1 = \frac{1}{m_2}\vec{P}_2 + [\underline{I}_2^{-1}(\vec{r}_2 \times \vec{P}_2)] \times \vec{r}_2 \tag{4}$$

These velocity vectors can then be transformed from their respective vehicle coordinate systems to a common contact plane coordinate system with axes n, t, z through the use of Euler angle transformation matrices $T_1$ and $T_2$ (the n-axis is normal to the contact plane while the t- and z-axes form an orthogonal set within the contact plane – see Figure 1). Using the transformation matrices, the post-impulse velocities at the impulse center (in n-t-z coordinates) are:

$$\underline{T}_1\vec{V}_{IC1f} - \underline{T}_1\vec{V}_{IC1i} = \frac{1}{m_1}\underline{T}_1\vec{P}_1 + \underline{T}_1([\underline{I}_1^{-1}(\vec{r}_1 \times \vec{P}_1)] \times \vec{r}_1)$$

$$\underline{T}_2\vec{V}_{IC2f} - \underline{T}_2\vec{V}_{IC2i} = \frac{1}{m_2}\underline{T}_2\vec{P}_2 + \underline{T}_2([\underline{I}_2^{-1}(\vec{r}_2 \times \vec{P}_2)] \times \vec{r}_2) \tag{5}$$

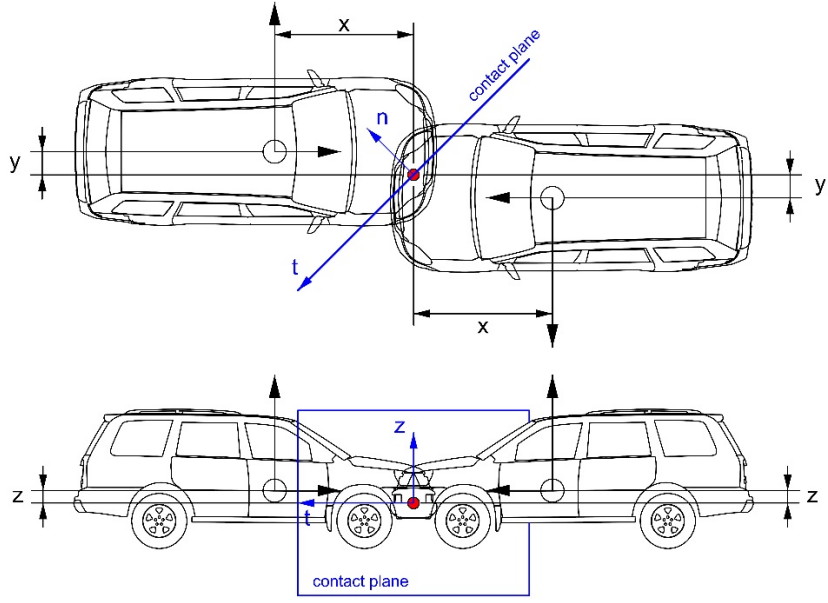

Figure 1: Impact Configuration Diagram

By subtracting the vehicle 1 equation from the vehicle 2 equation we arrive at the relative velocity vector (expressed in n-t-z coordinates) at the impulse center (for the sake of brevity the term "sliding velocity" will be used herein to refer to relative velocity at the impulse center along the contact plane):

$$\vec{V}_{relf} - \vec{V}_{reli} = \frac{1}{m_2}\underline{T}_2\vec{P}_2 - \frac{1}{m_1}\underline{T}_1\vec{P}_1 + \underline{T}_2\left(\left[\underline{I}_2^{-1}(\vec{r}_2 \times \vec{P}_2)\right] \times \vec{r}_2\right) - \underline{T}_1\left(\left[\underline{I}_1^{-1}(\vec{r}_1 \times \vec{P}_1)\right] \times \vec{r}_1\right) \quad (6)$$

Making use of Newton's third law, the vehicle impulses can be expressed as:

$$\underline{T}_1\vec{P}_1 = -\underline{T}_2\vec{P}_2 = \vec{P}$$

$$\vec{P}_1 = \underline{T}_1^{-1}\vec{P} = \underline{T}_1^T\vec{P} \qquad \vec{P}_2 = -\underline{T}_2^{-1}\vec{P} = -\underline{T}_2^T\vec{P} \quad (7)$$

Using these impulse relationships, the sliding velocity vector is found to be:

$$\vec{V}_{relf} - \vec{V}_{reli} = -\left[\left(\frac{1}{m_1} + \frac{1}{m_2}\right)\vec{P} + \underline{T}_1\left(\left[\underline{I}_1^{-1}(\vec{r}_1 \times \underline{T}_1^T\vec{P})\right] \times \vec{r}_1\right) + \underline{T}_2\left(\left[\underline{I}_2^{-1}(\vec{r}_2 \times \underline{T}_2^T\vec{P})\right] \times \vec{r}_2\right)\right] \quad (8)$$

The expanded version of this vector equation results in the following three important 3DIM equations relating the pre- and post-impulse sliding velocity components to the impulse components:

$$m'(V_{rnf} - V_{rni}) = -(a_1 P_n + b_1 P_t + c_1 P_z)$$

$$m'(V_{rtf} - V_{rti}) = -(a_2 P_n + b_2 P_t + c_2 P_z)$$

$$m'(V_{rzf} - V_{rzi}) = -(a_3 P_n + b_3 P_t + c_3 P_z) \quad (9)$$

$$m' = \frac{m_1 m_2}{m_1 + m_2}$$

The a, b, and c coefficients can be straightforwardly determined, though with substantial algebraic maneuvering, and contain vehicle mass/inertia terms, Euler angle transformation terms, and impact center location terms. The coefficients for the case of a vertically-oriented contact plane (normal axis is horizontal) and the t-axis oriented at some angle ϕ from the horizontal are provided in Appendix A.

## Three-Dimensional Impulse-Space

Before proceeding, it will be useful to discuss the visualization of impact mechanics solutions in impulse-space. This has been discussed in references [14] and [15] in the context of planar impact mechanics where fundamental lines of maximum compression and no-sliding were defined, and solutions relative to these lines were examined. In 3DIM impact configurations, the impulse space is characterized by fundamental planes derived from equations 10. Example fundamental planes are depicted in the left panel of Figure 2 and are the maximum compression plane, the t-axis no-sliding plane, and the z-axis no-sliding plane. Respectively, these planes represent solutions in which there is no relative velocity at the impulse center normal to the contact plane (i.e., zero restitution); no relative velocity in the t-axis direction; and no relative velocity in the z-axis direction. The point that lies at the intersection of these three fundamental planes is the maximum energy loss solution for a given impact configuration and is the impulse solution that results in all relative velocity components at the impulse center being brought to zero.

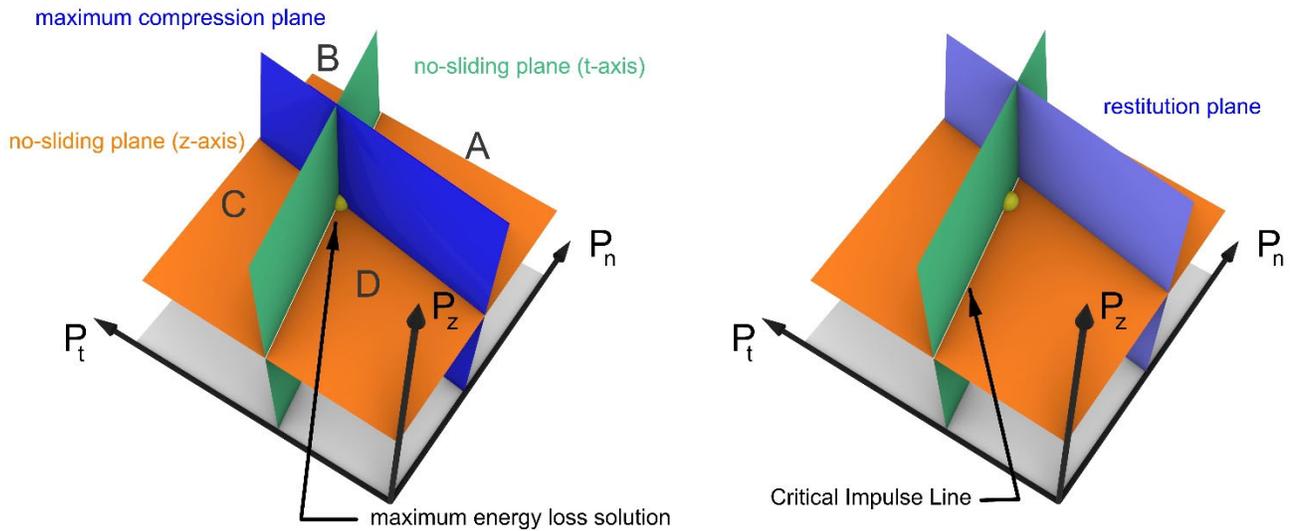

Figure 2: Three-Dimensional Impulse-Space Diagram

General solutions for the quadrant volumes labeled A, B, C, and D, in the left panel of Figure 2, are such that:

- Volume A: $P_n$ values are greater than the maximum compression plane, therefore, coefficient of restitution values are greater than zero; $P_t$ values are less than the t-axis no-sliding plane, therefore, continued sliding along the t-axis at the impulse center; $P_z$ values are greater than the z-axis no-sliding plane, therefore, reversed sliding velocity along the z-axis at the impulse center

- Volume B: $P_n$ values are greater than the maximum compression plane, therefore, coefficient of restitution values are greater than zero; $P_t$ values are greater than the t-axis no-sliding plane, therefore, reversed sliding along the t-axis at the impulse center; $P_z$ values are greater than the z-axis no-sliding plane, therefore, reversed sliding velocity along the z-axis at the impulse center
- Volume C: $P_n$ values are less than the maximum compression plane, therefore, coefficient of restitution values are less than zero; $P_t$ values are greater than the t-axis no-sliding plane, therefore, reversed sliding along the t-axis at the impulse center; $P_z$ values are greater than the z-axis no-sliding plane, therefore, reversed sliding velocity along the z-axis at the impulse center
- Volume D: $P_n$ values are less than the maximum compression plane, therefore, coefficient of restitution values are less than zero; $P_t$ values are less than the t-axis no-sliding plane, therefore, continued sliding along the t-axis at the impulse center; $P_z$ values are greater than the z-axis no-sliding plane, therefore, reversed sliding velocity along the z-axis at the impulse center

For the corresponding quadrant volumes below the z-axis no-sliding plane, the situation is the same except that, being below the no-sliding plane, continued sliding will occur along the z-axis at the impulse center for all cases.

In the right panel of Figure 2, the impulse-space is depicted with a restitution plane in place. For non-zero coefficient of restitution values, solutions will lie on a restitution plane that is oriented the same as the maximum compression plane but positioned along the normal axis based on the selected restitution value. One other fundamental feature of 3D impulse space is the intersection of the t- and z-axis no-sliding planes. Along this line of intersection is a locus of impulse solutions in which no post-impulse sliding velocity component along the contact plane exists at the end of the impulse. This line is fixed in impulse space for a given impact configuration and will herein be called the Critical Impulse Line.

## Contact Plane/Impulse Center Constraint Decisions

Returning to the impact mechanics equations derivation, and having arrived at equations 9, decisions need to be made regarding constraints applied at the impulse center. Through rigid-body kinematic relationships, we've reduced the 3DIM problem from a system of 12 equations and 15 unknowns (equations 1) to the three impulse center relative velocity equations above in which remain six unknowns (three impulse components and three post-impact impulse center relative velocity components). Three constraint relationships are yet still required. We'll start by extending the approach outlined for the PC-Crash 2D "Full-Impact" described in reference [12] to the three-dimensional case using a vertically-oriented contact plane with the normal and tangent axes aligned horizontally within the space-fixed reference system. The three constraints that make up the three-dimensional Full-Impact solution are that the three post-impact relative velocity components of equation 9 are all set to zero at the end of the compression phase of the impulse. A kinetic coefficient of restitution (Poisson's restitution) constraint is applied and the impulse solution for the Full-Impact condition is:

$$P_{nFI} = m'(1+\varepsilon)\frac{V_{rni}(b_2c_3 - b_3c_2) + V_{rti}(b_3c_1 - b_1c_3) + V_{rzi}(b_1c_2 - b_2c_1)}{a_1(b_2c_3 - b_3c_2) + b_1(a_3c_2 - a_2c_3) + c_1(a_2b_3 - a_3b_2)} \quad (10)$$

$$P_{tFI} = m'(1+\varepsilon)\frac{V_{rni}(a_3c_2 - a_2c_3) + V_{rti}(a_1c_3 - a_3c_1) + V_{rzi}(a_2c_1 - a_1c_2)}{a_1(b_2c_3 - b_3c_2) + b_1(a_3c_2 - a_2c_3) + c_1(a_2b_3 - a_3b_2)} \quad (11)$$

$$P_{zFI} = m'(1+\varepsilon)\frac{V_{rni}(a_2b_3 - a_3b_2) + V_{rti}(a_3b_1 - a_1b_3) + V_{rzi}(a_1b_2 - a_2b_1)}{a_1(b_2c_3 - b_3c_2) + b_1(a_3c_2 - a_2c_3) + c_1(a_2b_3 - a_3b_2)} \quad (12)$$

Note that the Full-Impact solution at zero restitution ($\varepsilon = 0$) is the maximum energy loss solution identified above in the impulse-space diagram of Figure 2.

Within the vertical contact plane (the t-z plane), the resultant impulse is:

$$P_{CP} = \sqrt{P_{tFI}^2 + P_{zFI}^2} \quad (13)$$

The Full-Impact solution holds for all user-selected "friction" ($\mu_{user}$) values greater than the following ratio:

$$\mu_{user} \geq \frac{\sqrt{P_{tFI}^2 + P_{zFI}^2}}{P_{nFI}} \quad (14)$$

For all user-selected values greater than this ratio PC-Crash reverts to the Full Impact solution. If selected value is less than this ratio, the impact is considered a Sliding Impact, in the PC-Crash vernacular, and impulse solutions are constrained to lie in a plane oriented in impulse space the same as that of the Full-Impact solution. The orientation of this plane is:

$$\phi_{FI} = \tan^{-1}\left(\frac{P_{zFI}}{P_{tFI}}\right) \quad (15)$$

An example $P_N$-$P_{CP}$ solution plane is shown in Figure 3 where, in the left panel, the solution plane is depicted in a perspective view in impulse space with respect to the fundamental planes and the Critical Impulse Line. In the right panel of this figure, the same configuration is depicted in a t-z plane view with the intersection of the t- and z-axis no-sliding planes and the $P_N$-$P_{CP}$ solution plane with the maximum compression plane also shown.

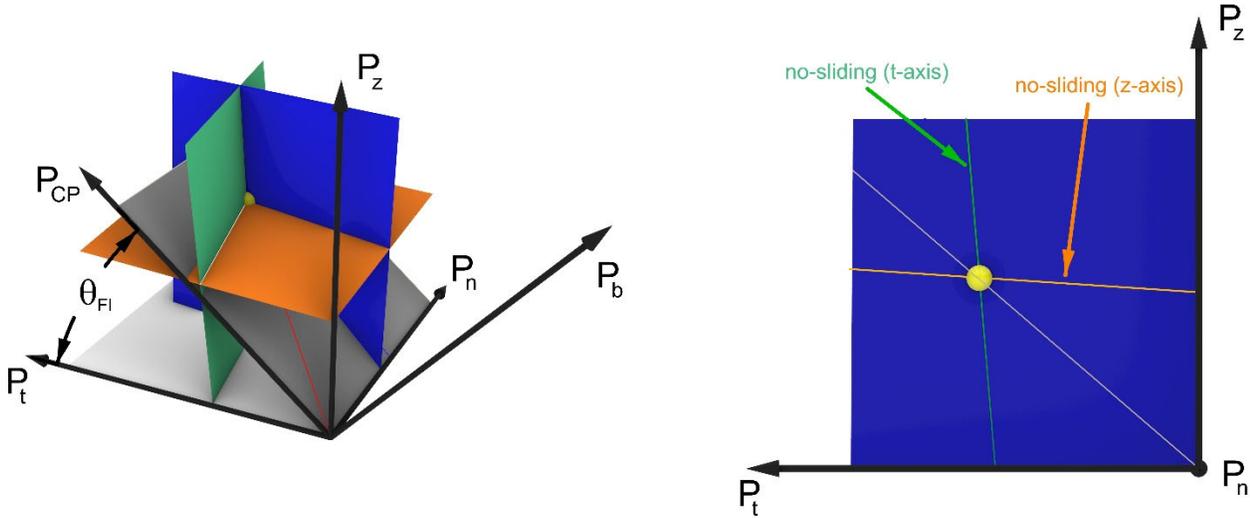

Figure 3. Full-Impact Contact Plane Configuration

The resultant contact plane impulse in the PC-Crash Sliding Impact scenario is further constrained by a relationship that takes the form of the Amonton-Coulomb friction law:

$$P_{CP} = \mu_{user} P_n \qquad (16)$$

While equation 16 is of the form of a friction law, and this term is referred to as a friction parameter by both PC-Crash and Virtual Crash, Brach [11] refers to this parameter as an impulse ratio and states emphatically that "The impulse ratio is *not* a coefficient of friction, although it can be related to it." He further states that this parameter "represents the retardation impulse that controls sliding along the tangential plane representing the crush surface." Given the complex interactions occurring between impacting automobiles at the contact surfaces, and the fact that the contact forces are resolved to occur at a single temporal and spatial point, this seems a reasonable interpretation and the term *impulse ratio* will be adopted for the remainder of this paper.

The governing impulse equations can be transformed to the $P_N$-$P_{CP}$ solution plane making it a pseudo-two-dimensional problem in that there is no impulse component perpendicular to this plane. The equations reduce to those of equations 17 with the a, b, and c coefficients determined for $\phi_{FI}$:

$$m'(V_{rnf} - V_{rni}) = -(a_1 P_n + b_1 P_{CP}) = -(a_1 + \mu_{user} b_1) P_n$$

$$m'(V_{rCPf} - V_{rCPi}) = -(a_2 P_n + b_2 P_{CP}) = -(a_2 + \mu_{user} b_2) P_n$$

$$m'(V_{rbf} - V_{rbi}) = -(a_3 P_n + b_3 P_{CP}) = -(a_3 + \mu_{user} b_3) P_n \qquad (17)$$

From the first of equations 17 above, the normal impulse, using either a kinematic or a kinetic coefficient of restitution, is found to be,

$$P_n = \frac{m'(1+\varepsilon) V_{rni}}{a_1 + \mu_{user} b_1} \qquad (18)$$

For two-dimensional planar impact mechanics configurations, Brach has discussed the condition where sliding velocity along the tangent axis comes to a stop at the end of the impulse, and has recommended this condition be used unless there is clear evidence of sliding throughout the impact [11]. The impulse ratio necessary to accomplish this condition is referred to by Brach as the Critical Impulse Ratio (CIR). For the pseudo-two-dimensional $P_N$-$P_{CP}$ plane, the CIR for a particular coefficient of restitution value can be found from the second of equations 17 for the condition where $V_{rCPf}$ is zero:

$$\mu_{CIR} = \frac{ra_1 - (1+\varepsilon)a_2}{(1+\varepsilon)b_2 - rb_1} \tag{19}$$

$$r = \frac{V_{rCPi}}{V_{rni}}; \quad V_{rCPi} = V_{rti}\cos\phi_{FI} + V_{rzi}\sin\phi_{FI}$$

The locus of CIR solutions in the $P_N$-$P_{CP}$ plane forms a no-sliding line in this plane (example shown in the right panel of Figure 4), but note here that this defines solutions that produce no post-impulse sliding velocity only along the $P_N$-$P_{CP}$ plane and, as the third of equations 17 indicates, does not preclude contact plane sliding velocity normal to the $P_N$-$P_{CP}$ plane. Generally speaking, the Critical Impulse Line will not necessarily lie in the $P_N$-$P_{CP}$ plane and, therefore, the only solution in this plane in which sliding velocity is fully brought to zero at the end of the impulse is the Full-Impact solution at maximum compression (i.e, $\varepsilon = 0$). We can define the intersection lines of the t- and z-axis no-sliding planes as shown in left panel of Figure 4 but, again, solutions in the $P_N$-$P_{CP}$ plane along these lines result only in no post-impact sliding velocity in those respective directions.

As discussed in references [15] and [16], the limiting impulse ratio of a PC-Crash Full-Impact coincides with that which Ishikawa defined as the Generalized Impulse Ratio (GIR), and is simply the critical impulse ratio at maximum compression. The GIR for the 3DIM Full-Impact solution is then.

$$\mu_{GIR} = \mu_{CIR}(\varepsilon = 0) = \frac{ra_1 - a_2}{b_2 - rb_1} \tag{20}$$

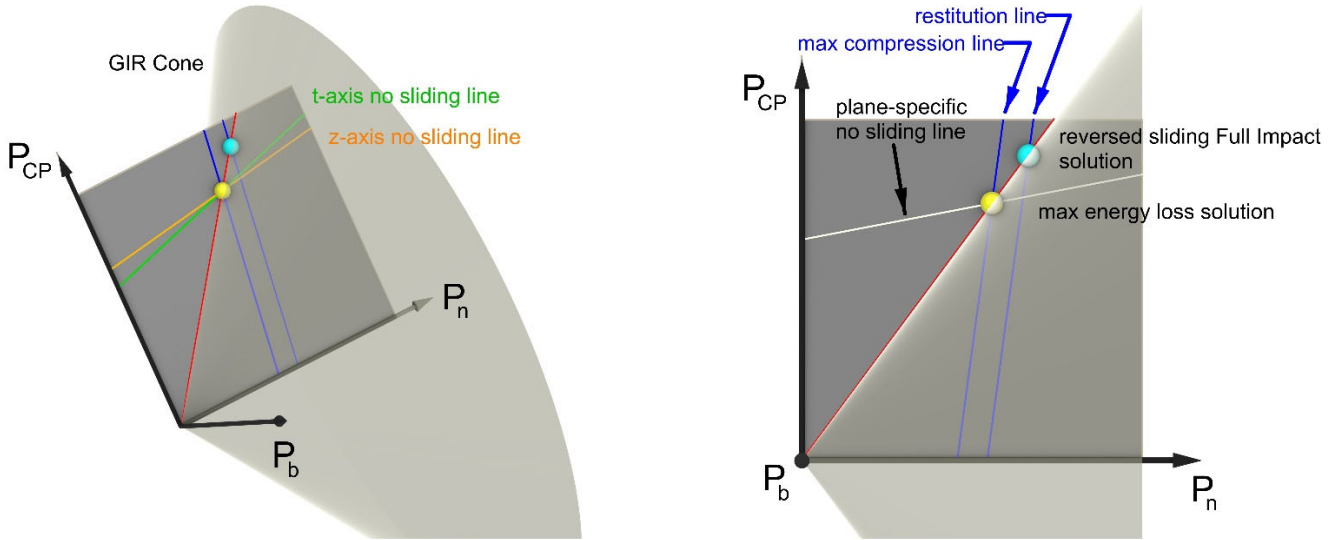

Figure 4. Impulse Space $P_N$-$P_{CP}$ Solution Plane and Full-Impact Friction Cone

In the left panel of Figure 4, the GIR cone is depicted in 3D-impulse space with the $P_N$-$P_{CP}$ solution plane. As noted above, in PC-Crash, any impulse ratio/friction value the analyst may wish to use outside the GIR cone simply reverts to the Full-Impact GIR solution. In doing so, the program prevents sliding velocity being brought to zero prior to maximum compression. If the user-selected impulse ratio value is less than the GIR, then the impulse cone narrows and the solutions are considered to be Sliding Impacts. As can be clearly seen in the right panel of Figure 4, a ramification of selecting the GIR as the limiting impulse ratio is that a user-defined impulse ratio/friction value can be selected that, for coefficient of restitution values greater than zero, place a so-called Sliding Impact solution in the Volume B impulse-space described above in Figure 2, and result in post-impact sliding velocities along the original t- and z-axes that are reversed from their pre-impact condition. Note also that, except for contact plane orientations that are perpendicular to the initial relative velocity vector, a sliding velocity along the contact plane will exist in the initial condition. A general conceptualization of the problem as one in which a static coefficient of friction must be overcome for sliding to commence is not apt as there is no static condition to begin with. Rather, given the pre-/post-impulse binary of an instantaneous impact-mechanics interaction, the relevant issue is what becomes of the initial contact plane sliding velocity in the post-impulse state. The contact plane impulse can (1) diminish it; (2) bring it to a stop; or (3) reverse its direction.

## Aligning with the Initial Sliding Velocity

Wach [17] presented a contact plane strategy in which the tangential axis orientation is set up such that the "tangent axis t coincides with the projection" of the pre-impact sliding velocity vector at the impulse center onto the contact plane. The contact plane angle for this strategy is:

$$\phi = \tan^{-1}\left(\frac{V_{rzi}}{V_{rti}}\right) \tag{21}$$

This then sets up the contact plane impulse to oppose the initial sliding velocity along the contact plane (this strategy is also discussed in the Virtual CRASH User Guide). A depiction of the contact plane configuration for this formulation is shown in Figure 5.

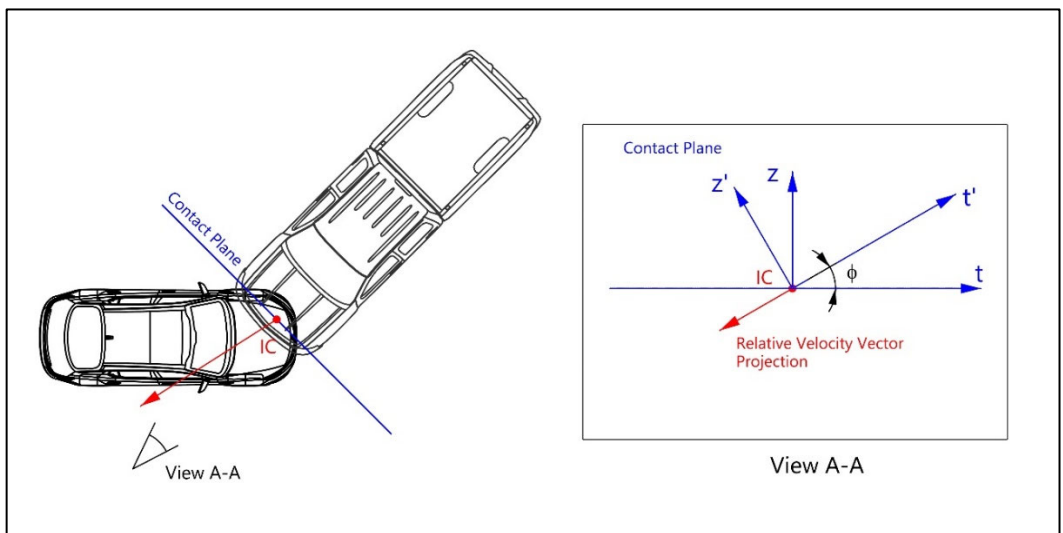

Figure 5: Wach's [17] Contact Plane Coordinate Configuration

As in the PC-Crash approach discussed above, the tangent axis impulse is defined by a user-selected impulse ratio parameter,

$$P_{t'} = \mu_{user} P_n \tag{22}$$

Wach employs Poisson's restitution coefficient and this, along with equation 22, comprise two constraints. A third constraint is yet needed and Wach provides that by imposing the condition that there is no impulse acting perpendicular to the n-t' plane (i.e., $P_{z'} = 0$).

For this approach, the solution proceeds as that of the Sliding Impact described above with the a, b, and c coefficients of equations 17 determined based on the angle ϕ. Wach limits impulse solutions to only those in which the impulse ratio is equal to or less than the $P_N$-$P_{t'}$ plane-specific Critical Impulse Ratio (equation 19 with plane-specific a, b, and c coefficients), thereby removing reversed in-plane sliding velocities from the solution set. The Virtual Crash User's Guide, on the other hand, defines the limit to their solutions vis-à-vis the $P_N$-$P_{t'}$ plane-specific Generalized Impulse ratio. Both of these limiting solutions are shown in Figure 7. An important item to note here is that the angles ϕ and ϕ$_{FI}$ will not generally coincide and, therefore, the solution planes for Wach's method and that of PC-Crash will likely not be the same. This has the important ramification that Wach's method, unless ϕ and ϕ$_{FI}$ happen to be the same, cannot replicate a true Full-Impact solution and, unless the solution happens to coincide with the Critical Impulse Line, will therefore produce out-of-plane sliding velocity. An out-of-plane directional change of the sliding velocity vector has been referred to as "swerve" by Stronge [18] and, within this current formulation, except for a Full Impact solution, is out of the direct control of the analyst as there is no constraint in place to deal with it.

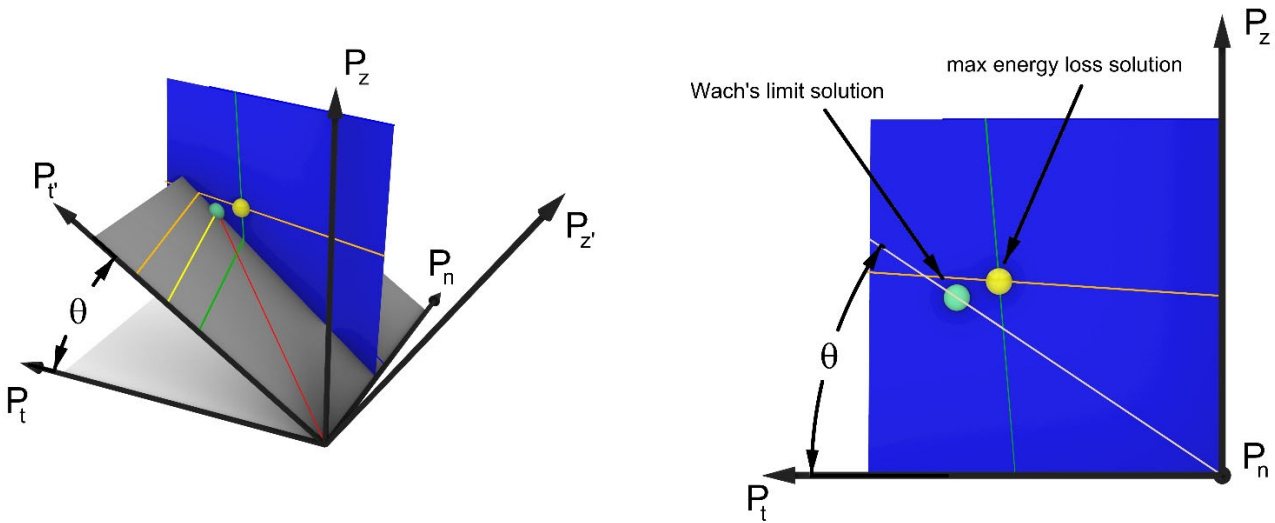

Figure 6: Wach Impulse Solution Plane Configuration

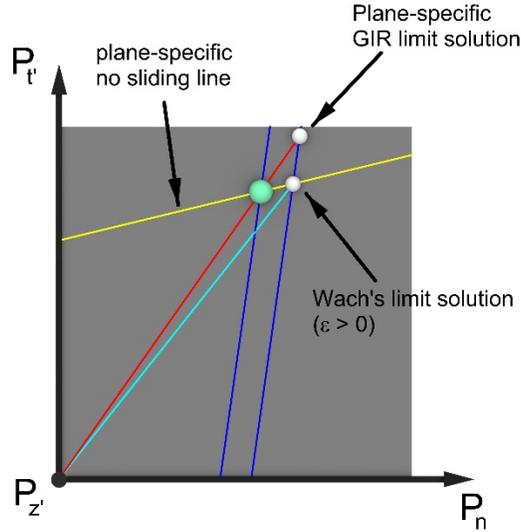

Figure 7: Limiting Impulse Ratios on the Wach Solution Plane

While the Virtual Crash User's Guide describes the use of a 3DIM approach following that of Wach's, in practice, this is not what that program appears to do. In fact, it seems to use the 3D PC-Crash approach described above. To demonstrate this, a simple 90-degree side impact configuration was set up in Virtual Crash. Figure 8 depicts this impact configuration in which the front of a Chevrolet Suburban traveling 25 mph strikes the right side of a Chevrolet Malibu traveling 25 mph with no vertical components to either of these initial velocities. Both vehicles are were available in the Virtual Crash vehicle database and no changes to them were made. The contact plane was automatically setup by Virtual Crash with the normal axis oriented 90 degrees relative to the space-fixed coordinate system. In the upper panel of this figure, the configuration is set up with an impulse ratio/friction value of 1.0 (Virtual Crash default value) and the impact is a Full Impact as is evidenced by the impulse vector placed within the periphery of the impulse cone. In the lower panel, the impulse ratio/friction value is 0.4 and the impact is a Sliding Impact as is evident by the impulse vector being located at the periphery of the narrowed cone. Other than the difference in friction values the two example impact configurations are identical. The initial relative velocity vector in these configurations is purely in the x-y (horizontal) plane yet both solutions result in a vertical component for the resulting impulse (i.e., the nz angles are non-zero). Using the stated impulse vector orientation values (ni and nz) highlighted in Figure 8 it can be shown, using equation 29, that both solutions result in the same contact plane impulse ratio ($P_z/P_t$). This then indicates both solutions reside on the Full Impact impulse-space solution plane and adhere to the PC-Crash approach described above.

$$\frac{P_z}{P_t} = \frac{\tan n_z}{\sin(n_i - \psi_o)} \qquad (29)$$

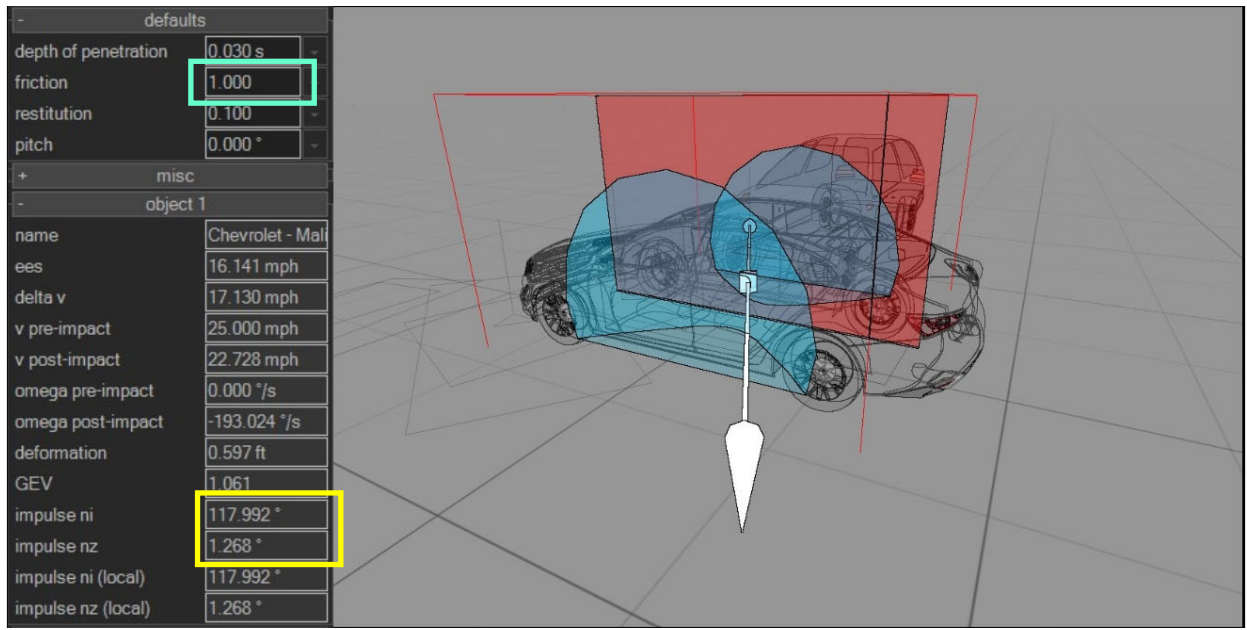

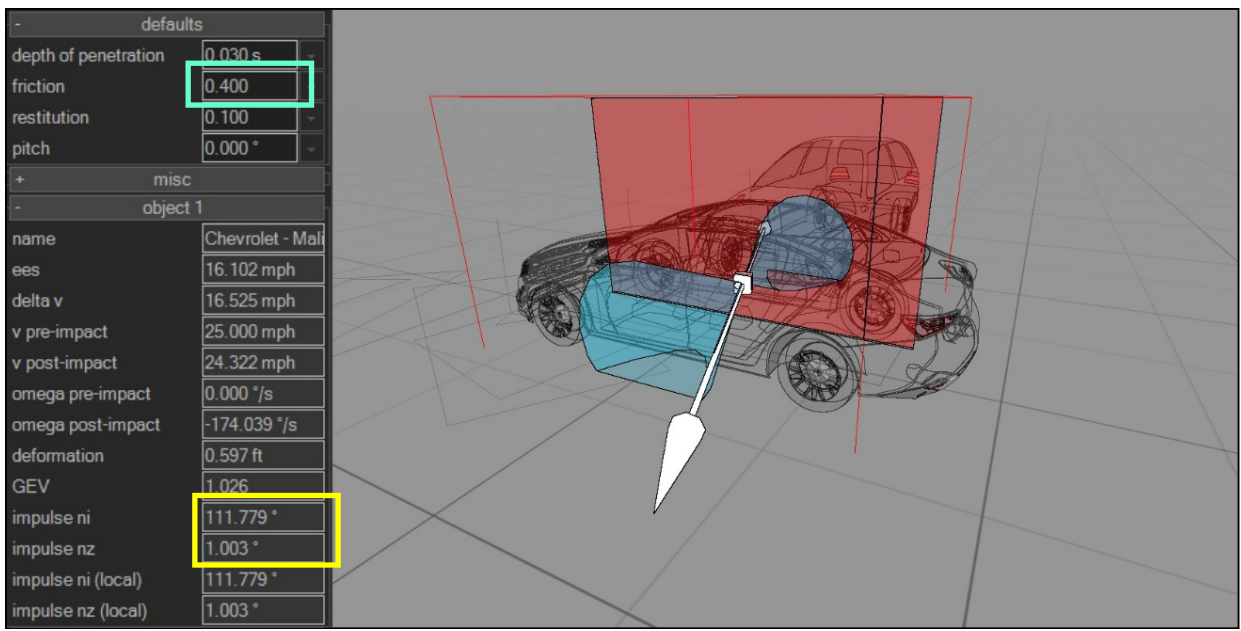

Figure 8: Virtual Crash Impulse Plane Example

The initial sliding velocity in both configurations was 25 mph in the positive direction of the horizontal tangential axis (rearward relative to the Malibu for a contact plane normal-axis orientation of 90 degrees). The post-impact sliding velocity along the tangential axis was -2.2 mph (reversed sliding) for the Full Impact configuration and +3.5 mph (continued sliding) for the Sliding Impact configuration. Both example configurations resulted in small swerve components of less than 0.5 mph; downward relative to the Malibu for the Full Impact configuration and upward for the Sliding Impact configuration. The example Sliding Impact configuration is essentially setup as described by Wach (and the Virtual Crash User's Guide) but nonetheless results in vertical impulse and swerve velocity components.

## Independent Impulse Ratios in the Contact Plane

Except for the Full Impact solution, the 3DIM approaches thus far examined have been shown to result in post-impact sliding velocity components in the contact plane that are not in the control of the analyst. An approach in which this control can be gained is that discussed by Brach in reference [10] where independent impulse ratios for the orthogonal impulse components in the contact plane are used. In working with this approach, as with those discussed above, a restitution constraint is used in the normal direction. The necessary second and third constraints are user-selected impulse ratios applied to the orthogonal t- and z-axes impulses. The relative velocity equations for this approach are then:

$$m'(V_{rnf} - V_{rni}) = -(a_1 + b_1\mu_t + c_1\mu_z)P_n$$
$$m'(V_{rtf} - V_{rti}) = -(a_2 + b_2\mu_t + c_2\mu_z)P_n$$
$$m'(V_{rzf} - V_{rzi}) = -(a_3 + b_3\mu_t + c_3\mu_z)P_n \tag{23}$$

From the first of these equations, the normal impulse is readily found and the overall impulse solution is:

$$\vec{P} = \begin{bmatrix} 1 \\ \mu_t \\ \mu_z \end{bmatrix} P_n \quad where \quad P_n = \frac{m'(1+\varepsilon)V_{rni}}{a_1 + \mu_t b_1 + \mu_z c_1} \tag{24}$$

For the case where there is no post-impact sliding velocity in the t-direction at the impulse center, the corresponding t-axis critical impulse ratio, derived from the second of equations 23, is:

$$\mu_{tC} = \eta_t \mu_z + \mu_{tC0}$$

$$\eta_t = \frac{rc_1 - (1+\varepsilon)c_2}{(1+\varepsilon)b_2 - rb_1}; \quad \mu_{tC0} = \frac{ra_1 - (1+\varepsilon)a_2}{(1+\varepsilon)b_2 - rb_1} \tag{25}$$

Similarly, using the third of equations 23, a critical impulse ratio for the condition where the z-direction post-impact sliding velocity is brought to a stop at the end of the impulse is:

$$\mu_{zC} = \eta_z \mu_t + \mu_{zC0}$$

$$\eta_z = \frac{r_z b_1 - (1+\varepsilon)b_3}{(1+\varepsilon)c_3 - r_z c_1}; \quad \mu_{zC0} = \frac{r_z a_1 - (1+\varepsilon)a_3}{(1+\varepsilon)c_3 - r_z c_1}; \quad r_z = \frac{V_{rzi}}{V_{rni}} \tag{26}$$

For the special case where there is no post-impact sliding velocity at all (i.e., where the Critical Impulse Line intersects the particular restitution plane), the respective critical impulse ratio values become:

$$\mu_{tC} = \frac{\eta_t \mu_{zC0} + \mu_{tC0}}{1 - \eta_t \eta_z}; \quad \mu_{zC} = \frac{\eta_z \mu_{tC0} + \mu_{zC0}}{1 - \eta_t \eta_z} \tag{27}$$

The use of independent impulse ratios would seem to have some relevance for automobile collision applications. The aggregate interaction of deforming, irregular surfaces at the crush interface between colliding vehicles shouldn't be expected to respond like one clearly isotropic surface striking another. In this context, there is no reason to assume that the impulse ratios should necessarily be the same in all directions of a selected contact plane. Of course, the question regarding what these values should be for a given impact configuration immediately arises - these are not tabulated friction values that can be readily looked up after all. While, the independent impulse ratio formulation presented above provides the analyst with more control of the contact plane sliding velocity, it requires the specification of two impulse ratio values and the additional engineering judgement that that necessitates. Correspondingly, this approach has not found popular use as an accident reconstruction tool.

## Conclusion

For a given impact configuration, there exists a line in impulse space that represents a locus of solutions in which a sliding velocity within the contact plane at the end of the impulse will be zero. This is called the Critical Impulse Line and impulse solutions that do not fall on that line result in a post-impulse sliding velocity, in some direction, within the contact plane. This can include directions that differ from the initial sliding velocity; this is known as a swerve.

A PC-Crash Full-Impact solution, with the coefficient of restitution value equal to zero, lies at a point in impulse space that is on the Critical Impulse Line and coincides with the point of maximum energy loss.

A PC-Crash Full-Impact solution with a non-zero coefficient of restitution will reside on a plane defined by the contact plane components of the zero restitution solution (equation 15) and result in a post-impulse sliding velocity that will be in the opposite direction to that of the initial sliding velocity. The PC-Crash Sliding Impact solutions are also constrained to lie in this plane.

Wach has presented an approach in which an impulse-space plane is set up such that the contact plane impulse opposes the initial relative velocity component that is perpendicular to the established normal direction. Solutions using this approach will generally lie in an impulse-space plane in which a Full Impact solution cannot be found and one in which out-of-plane post-impact velocities will generally always exist, even for Wach's limiting critical impulse ratio.

While the Virtual Crash User's Guide indicates that their impact model follows Wach's approach (albeit with a different limiting impulse ratio), in practice it appears instead to follow the PC-Crash approach.

A rigid-body impact model in which the user can be in direct control of the contact plane sliding velocity components has been discussed by Brach and has been briefly reviewed here. Ultimately, this model provides the user the ability to specify impulse solutions along the Critical Impulse Line if one so chooses. A practical drawback to this method is the additional engineering judgement necessary in the specification of two independent impulse ratio values.

Energy loss considerations have not been discussed here. It will simply be noted that an algebraic, rigid-body impact model is, as Goldsmith put it [19], "…incapable of describing the transient stresses, forces, or deformations produced…" While net changes in kinetic energy can be readily determined and the "contributions" of the involved impulses fleshed out, algebraic impact models based on rigid-body assumptions simply do not possess the sophistication to further discern energy loss due to specific physical mechanisms occurring along the contact plane. For the interested reader, references [14] and [20] provide useful discussions on rigid-body impact mechanics energy loss.

## Appendix A: Coefficient Definitions

In an effort achieve some brevity for the resulting coefficient definitions, vehicle symmetry about the x-z plane will be assumed. This means the product of inertia terms $I_{xy}$ and $I_{yz}$ are zero. The $I_{xz}$ product of inertia term is typically small compared to the primary inertia terms and, as it is in PC-Crash and Virtual Crash, will here be considered sufficiently small to neglect. The inertia tensors are then reduced to:

$$\underline{I_1} = \begin{bmatrix} I_{x1} & 0 & 0 \\ 0 & I_{y1} & 0 \\ 0 & 0 & I_{z1} \end{bmatrix} \qquad \underline{I_2} = \begin{bmatrix} I_{x2} & 0 & 0 \\ 0 & I_{y2} & 0 \\ 0 & 0 & I_{z2} \end{bmatrix}$$

A coordinate transformation matrix from a vehicle coordinate system to a contact plane coordinate system with the following rotation sequence will be used: heading angle ($\gamma$), attitude angle ($\theta$), and roll angle ($\phi$):

$$\underline{T} = \begin{bmatrix} \cos\gamma\cos\theta & \sin\gamma\cos\theta & -\sin\theta \\ -\sin\gamma\cos\phi + \cos\gamma\sin\theta\sin\phi & \cos\gamma\cos\phi + \sin\gamma\sin\theta\sin\phi & \cos\theta\sin\phi \\ \sin\gamma\sin\phi + \cos\gamma\sin\theta\cos\phi & -\cos\gamma\sin\phi + \sin\gamma\sin\theta\cos\phi & \cos\theta\cos\phi \end{bmatrix}$$

With this transformation matrix, the a, b, and c coefficients for an impact configuration in which $\theta$ is zero (i.e., the normal axis is horizontal) are:

$a_1 = A + m'f_1$

$a_2 = -B\cos\phi + m'(f_2\cos\phi + f_3\sin\phi)$

$a_3 = m'f_3\cos\phi + m'(f_7 - f_2)\sin\phi$

$b_1 = a_2$

$b_2 = (1 + C\cos^2\phi) + m'(f_4\cos^2\phi + 2f_5\sin\phi\cos\phi + f_6\sin^2\phi)$

$b_3 = m'(f_5(\cos^2\phi - \sin^2\phi) + (f_8 - f_4 + f_6)\sin\phi\cos\phi)$

$c_1 = a_3$

$c_2 = b_3$

$c_3 = 1 + m'(f_6\cos^2\phi + (f_4 + f_8)\sin^2\phi - 2f_5\sin\phi\cos\phi)$

Where the coefficients A, B, and C are those of Brach's planar derivation [10] and are:

$$A = 1 + m'\left[\frac{(-x_1\sin\gamma_1 + y_1\cos\gamma_1)^2}{I_{z1}} + \frac{(-x_2\sin\gamma_2 + y_2\cos\gamma_2)^2}{I_{z2}}\right]$$

$$B = m'\left[\frac{(x_1\cos\gamma_1 + y_1\sin\gamma_1)(-x_1\sin\gamma_1 + y_1\cos\gamma_1)}{I_{z1}} + \frac{(x_2\cos\gamma_2 + y_2\sin\gamma_2)(-x_2\sin\gamma_2 + y_2\cos\gamma_2)}{I_{z2}}\right]$$

$$C = m' \left[ \frac{(x_1 \cos \gamma_1 + y_1 \sin \gamma_1)^2}{I_{z1}} + \frac{(x_2 \cos \gamma_2 + y_2 \sin \gamma_2)^2}{I_{z2}} \right]$$

The coefficients $f_1$ through $f_8$ are:

$$f_1 = z_1^2 \left[ \frac{\sin^2 \gamma_1}{I_{x1}} + \frac{\cos^2 \gamma_1}{I_{y1}} \right] + z_2^2 \left[ \frac{\sin^2 \gamma_2}{I_{x2}} + \frac{\cos^2 \gamma_2}{I_{y2}} \right]$$

$$f_2 = z_1^2 \sin \gamma_1 \cos \gamma_1 \left( \frac{1}{I_{x1}} - \frac{1}{I_{y1}} \right) + z_2^2 \sin \gamma_2 \cos \gamma_2 \left( \frac{1}{I_{x2}} - \frac{1}{I_{y2}} \right)$$

$$f_3 = - \left[ z_1 \left( \frac{x_1 \cos \gamma_1}{I_{y1}} + \frac{y_1 \sin \gamma_1}{I_{x1}} \right) + z_2 \left( \frac{x_2 \cos \gamma_2}{I_{y2}} + \frac{y_2 \sin \gamma_2}{I_{x2}} \right) \right]$$

$$f_4 = z_1^2 \left( \frac{\cos^2 \gamma_1}{I_{x1}} + \frac{\sin^2 \gamma_1}{I_{y1}} \right) + z_2^2 \left( \frac{\cos^2 \gamma_2}{I_{x2}} + \frac{\sin^2 \gamma_2}{I_{y2}} \right)$$

$$f_5 = z_1 \left( \frac{x_1 \sin \gamma_1}{I_{y1}} - \frac{y_1 \cos \gamma_1}{I_{x1}} \right) + z_2 \left( \frac{x_2 \sin \gamma_2}{I_{y2}} - \frac{y_2 \cos \gamma_2}{I_{x2}} \right)$$

$$f_6 = \left( \frac{y_1^2}{I_{x1}} + \frac{x_1^2}{I_{y1}} \right) + \left( \frac{y_2^2}{I_{x2}} + \frac{x_2^2}{I_{y2}} \right)$$

$$f_7 = \frac{(y_1^2 - x_1^2) \sin \gamma_1 \cos \gamma_1 + x_1 y_1 (\cos^2 \gamma_1 - \sin^2 \gamma_1)}{I_{z1}}$$
$$+ \frac{(y_2^2 - x_2^2) \sin \gamma_2 \cos \gamma_2 + x_2 y_2 (\cos^2 \gamma_2 - \sin^2 \gamma_2)}{I_{z2}}$$

$$f_8 = - \left[ \frac{(x_1 \cos \gamma_1 + y_1 \sin \gamma_1)^2}{I_{z1}} + \frac{(x_2 \cos \gamma_2 + y_2 \sin \gamma_2)^2}{I_{z2}} \right]$$